\documentclass[prx,twocolumn,notitlepage,aps,showpacs,longbibliography,amsmath,amstex,amssymb,mathfonts]{revtex4-1}

\pdfoutput=1

\usepackage[pdftex]{graphicx}

\usepackage{amsmath}
\usepackage{amssymb}
\usepackage{mathtools}
\usepackage{verbatim}
\usepackage{hhline}

\usepackage[protrusion=true,expansion=true]{microtype}

\usepackage{hyperref}
\hypersetup{
	colorlinks=true,       		
	linkcolor=black,          	
	citecolor=blue,        		
	urlcolor=blue          		
}

\usepackage{color}
\usepackage{array}
\usepackage{bbm}
\usepackage{hyperref}
\usepackage{color}
\usepackage{array}
\usepackage{cancel} 
\usepackage{ulem} 
\usepackage{slashed}

\usepackage{booktabs}
\newcolumntype{C}[1]{>{\centering\arraybackslash}m{#1}}
\AtBeginDocument{
\heavyrulewidth=.08em
\lightrulewidth=.05em
\cmidrulewidth=.03em
\belowrulesep=.65ex
\belowbottomsep=0pt
\aboverulesep=.4ex
\abovetopsep=0pt
\cmidrulesep=\doublerulesep
\cmidrulekern=.5em
\defaultaddspace=.5em
}
\usepackage{booktabs}
\newcolumntype{R}[1]{>{\raggedleft\arraybackslash}p{#1}}
\AtBeginDocument{
\heavyrulewidth=.08em
\lightrulewidth=.05em
\cmidrulewidth=.03em
\belowrulesep=.65ex
\belowbottomsep=0pt
\aboverulesep=.4ex
\abovetopsep=0pt
\cmidrulesep=\doublerulesep
\cmidrulekern=.5em
\defaultaddspace=.5em
}

\newcommand{\<}{\langle}

\newcommand{\up}{\uparrow}
\newcommand{\down}{\downarrow}
\renewcommand{\>}{\rangle}
\renewcommand{\(}{\left(}
\renewcommand{\)}{\right)}
\renewcommand{\[}{\left[}
\renewcommand{\]}{\right]}
\renewcommand{\v}[1]{\mathbf{#1}} 

\renewcommand{\d}{\partial}

\newcommand{\eps}{\epsilon}
\newcommand{\p}{\parallel}
\renewcommand{\t}{\tau}

\begin{document}

\title{Symmetry enforced fractonicity and $2d$ quantum crystal melting}

\author{Ajesh Kumar}
\affiliation{Department of Physics, University of Texas at Austin, Austin, Texas 78712, USA}
\author{Andrew C. Potter}
\affiliation{Department of Physics, University of Texas at Austin, Austin, Texas 78712, USA}

\begin{abstract}
Fractons are particles that cannot move in one or more directions without paying energy proportional to their displacement. Here, we introduce the concept of symmetry enforced fractonicity, in which particles are fractons in the presence of a global symmetry, but are free to move in its absence. A simple example is dislocation defects in a two-dimensional crystal, which are restricted to move only along their Burgers vector due to particle number conservation. Utilizing a recently developed dual rank-2 tensor gauge description of elasticity, we show that accounting for the symmetry enforced one-dimensional nature of dislocation motion dramatically alters the structure of quantum crystal melting phase transitions. We show that, at zero temperature, sufficiently strong quantum fluctuations of the crystal lattice favor the formation of a super-solid phase that spontaneously breaks the symmetry enforcing fractonicity of defects. The defects can then condense to drive the crystal into a super-nematic phase via a phase transition in the $2+1d$ XY universality class to drive a melting phase transition of the crystal to a nematic phase. This scenario contrasts the standard Halperin-Nelson scenario for thermal melting of $2d$ solids in which dislocations can proliferate via a single continuous thermal phase transition. We comment on the application of these results to other scenarios such as vortex lattice melting at a magnetic field induced superconductor-insulator transition, and quantum melting of charge density waves of stripes in a metal.

\end{abstract}
\maketitle

\section{Introduction}
Condensation of topological defects plays a crucial role in the thermal and quantum melting of symmetry breaking orders in low dimensional systems. Famous examples include the destruction of $2d$ superfluids or easy-plane magnets by vortex proliferation (the BKT transition~\cite{berezinskii1971destruction,berezinskii1972destruction,kosterlitz1973ordering}). The critical properties of these transitions are captured by the $2d$ XY universality class at finite temperature, and extend to a related zero-temperature quantum phase transition that is in the $3d$ XY universality class, with the extra dimension encoding the quantum dynamics and fluctuations of the vortices.

The melting of $2d$ crystalline solids provides another classic example of destruction of order by topological defects. At non-zero temperature, $2d$ crystals exhibit dislocation defects, whose thermal proliferation can drive a continuous melting transition into a nematic or hexatic phase in which continuous translation symmetry is restored, but rotational symmetry breaking persists. As described by textbook Halperin-Nelson theory~\cite{halperin1978theory, nelson1979dislocation, young1979melting}, this thermal dislocation-induced melting is essentially the same as the vortex condensation in a superfluid, with the minor distinction that the dislocations come in two flavors distinguished by different Burgers vectors. 
In the spirit of the vortex melting of superfluids, an analogous zero-temperature quantum melting transition in the $3d$ XY universality class was hypothesized, and invoked in various theories of melting of electronic crystals such as charge- and spin- density waves~\cite{kivelson1998electronic,fradkin1999liqcrystalphases}, and stripes in high-temperature cuprate superconducting compounds~\cite{zaanen2001geometric,zhang2002competingorders,sachdev2002stronglycoupled,nussinov2002stripe,mross2012theory,mross2012stripe}, and neutral atomic crystals~\cite{radzihovsky2011fluctuations}. 

However, whereas the thermal $2d$ solid melting transition occurs via the entropic proliferation of defect configurations, a putative zero-temperature quantum dislocation condensation transition must occur via the virtual (tunneling) dynamics of dislocations. Here, the analogy between vortices and dislocations breaks down, due to strong symmetry-constraints that restrict the dynamics of dislocations. Namely, while dislocations may freely move (or ``glide") along their Burgers vector, they cannot ``climb" perpendicular to this direction without adding or removing particles from the system. This so-called ``glide-constraint" has been previously been noted in the literature~\cite{hirth1982theory,marchetti1999interstitials,cvetovic2006topological}, but, as we will argue, its consequences for quantum crystal melting have not been fully appreciated. Specifically, in an insulating crystal made of particles with a conserved number, changing the particle number costs a non-zero amount of energy, and cannot occur at zero-temperature. Similarly, since the condensation of dislocations also requires a condensation of these symmetry-forbidden climb events, producing a quantum superposition of states with different particle numbers, i.e. a dislocation condensate is necessarily accompanied by superfluid order.

In other words, a direct quantum melting transition via dislocation condensation would necessarily involve simultaneous restoration of translation symmetry and breaking of particle-number conservation. In the conventional Landau paradigm of phase transitions, a direct transition between two unrelated symmetry breaking patterns is generically not possible, and requires either fine-tuning or the emergence of exotic deconfined particles and gauge fields~\cite{senthil2004deconfined,senthil2004deconfined2}. Therefore, conventional wisdom would dictate that the climb constraint forbids a continuous (second order) phase transition driven by quantum dislocation condensation.

To address this point, we construct an effective field theory of dislocations, utilizing a dual description of the elastic fluctuations of the crystal~\cite{zaanen2004duality}, which was recently and elegantly reformulated as a higher rank gauge theory~\cite{pretko2018fracton}. In the latter formulation, the dual gauge charges are disclinations, that cannot move without exciting a finite number of additional excitations per unit displacement~\cite{harris1971intrinsic,dewit1971relation}, and thus at zero-temperature are immobile objects, dubbed ``fractons"~\footnote{Strictly speaking, the term fractons applies to immobile objects created by fractal shaped operators, such as those in Haah's code~\cite{haah2011code}. However, following recent literature, we will use this term to refer to any particle that is immobile along one more more spatial directions.}. 

The gauge charges of this theory are disclinations (orientational defects), and are completely immobile fractonic objects. The dislocations of the crystal appear in this dual description as dipoles of the gauge charge, and are not inherently fractonic. Namely, in the absence of any symmetries, they can move in any direction without producing additional excitations. However, we will show that the glide-constraint dictates that, in the presence of $U(1)$ particle-number conservation symmetry, the dislocations cannot hop perpendicular to their Burgers vector without producing excitations with a net $U(1)$ charge. In a charge insulating crystal, these are gapped and hence the symmetry forces the dislocations to be $1d$ sub-dimensional ``fracton" particles. We dub this phenomena: symmetry enforced fractonicity.

This symmetry enforced fractonicity rules out the existence of an insulating nematic phase, as condensation of dislocations inevitably produces condensation of particle number -- i.e producing a superfluid. Moreover, it implies that a conventional, continuous quantum phase transition between a charge-insulating crystal and any non-insulating nematic or hexatic phase is fundamentally forbidden. Instead, we show that, rather than directly melting the crystal, quantum fluctuations of the crystal can instead drive condensation of the underlying particles to produce a super-solid phase with coexisting superfluid and crystalline orders. In this super-solid, the superfluid condensate alleviates the symmetry-enforced fractonicity, and frees the dislocations to move in any direction. As quantum fluctuations of the crystal order are further increased, it is then possible to follow a quantum analog of the Halperin-Nelson theory, in which the super-solid directly transitions into a super-nematic or super-hexatic, before ultimately condensing the disclination defects to completely restore the translation symmetry and arrive at a superfluid phase. 

After describing this sequence of transitions using the dual higher-rank gauge theory language, we next explore the consequences for these ideas to other types of crystals. We first adapt these ideas to the quantum melting of vortex lattices in superfluids or superconductors. Finally, we explore the consequences of symmetry-enforced fractonicity in the quantum melting of $2d$ charge-density waves or ``stripes" in a metal, where our analysis suggests that quantum fluctuations of stripe dislocations can favor pairing and superconductivity.

Before embarking on the main subject of this paper, we briefly comment on the relation of our results to those previously obtained in the literature. Having mapped $2+1d$ elasticity theory to a higher rank tensor gauge theory, Pretko and Radzihovsky~\cite{pretko2018fracton} describe the Halperin-Nelson sequence of melting transitions in the gauge theory language, and further predict the possibility of a supersolid phase at zero temperature, which agrees with our analysis. We work out the zero temperature phase diagram of the dual gauge theory, and show that quantum melting occurs through an intermediate supersolid phase. We also discuss the finite temperature phase diagram in Sec.~\ref{subsec:finite_temperature}, and in particular, recover the Halperin-Nelson melting scenario.

A dual gauge-theory of elasticity was first formulated in works by Zaanen \textit{et al}.~\cite{zaanen2004duality} reviewed in Ref.~\cite{beekman2017dual}, which then studied quantum melting transitions of a $2d$ crystal. 
These works consider the glide constraint, and also identify that a dislocation condensate has coexisting nematic and superfluid order. However, these works primarily neglect the role of excitations with non-zero particle number (interstitials and vacancies), and their role in the origins of the superfluid. Moreover, they posit that a $2d$ crystal can melt via a single continuous transition to a smectic superfluid in the $3d$ XY universality class.
In contrast, we explicitly incorporate interstitials and their coupling to the dislocation motion, and find that the onset of a superfluid of interstitials and vacancies, and hence the resulting alleviation of the symmetry enforced fractonicity of the dislocations, is crucial for the disordering of crystalline order, and argue that the quantum nematic phase is likely preceded by an intermediate super-solid phase.

There are also previous works studying thermal vortex-line lattice melting in $3d$~\cite{marchetti1990dislocation,marchetti1999interstitials}. Building on the work of Marchetti and Nelson~\cite{marchetti1990dislocation}, where they describe the melting by an unbinding transition of dislocation loops, Marchetti and Radzihovsky~\cite{marchetti1999interstitials} incorporate the coupling of the climb motion of dislocation loops out of their glide plane to interstitials and vacancies. They also argue that melting of a vortex solid to a vortex hexatic phase would require an intermediate vortex supersolid phase if the transitions are continuous, in agreement with our discussion on $2+1d$ quantum vortex lattice melting in Sec.~\ref{sec:vortex_lattice}.

Looking beyond elasticity, our definition of symmetry enforced fractonicity is conceptually related to recently proposed ideas of fractal-symmetry protected topological phases~\cite{devakul2018fractal}. In contrast, here, we will consider only ordinary global symmetries, rather than a more exotic infinite number of symmetries defined on different fractal-geometry sub-systems.

\section{Symmetry enforced fractonicity}
We first investigate the quantum melting of a $2d$ solid formed by a crystal of bosonic atoms with a conserved number. To formulate an effective field theory description of symmetry enforced fractonicity of dislocations, we begin with the standard theory of elastic fluctuations of a crystal in terms of a displacement field $u_i(\v{r},\tau)$, that describes the displacement the $i\in \{x,y\}$ direction of the atom located from its equilibrium position $\vec{r}$, at imaginary time $\tau$ (this Euclidean time coordinate is related to the real-time by the usual Wick rotation: $\tau = it$). Here, we adopt a continuum description, obtained by coarse-graining on a length-scale that includes many unit cells of the underlying crystal lattice. In this limit, we may approximate the discrete atomic density by continuous density and current fields: 
\begin{align}
j_\mu(x) &=\begin{pmatrix} -i\rho \\ \v{j} \end{pmatrix} \approx 
(-i)\begin{pmatrix} 
\rho_0\(1-\nabla\cdot \v{u}\)\\
\d_\tau u_i
\end{pmatrix} 
+\mathcal{O}(\nabla^2u).
\end{align}
Here, $x_\mu = \begin{pmatrix}\tau \\ \vec{r}\end{pmatrix}$ is the Euclidean space-time coordinate, and $\rho_0$ is the average density of particles in the crystal. For smooth elastic fluctuations of atomic displacements, $\v{u}^s$, these currents satisfy the continuity equation: $\d_\mu j_\mu = i\[\d_\tau\(\nabla \cdot\v{u}^s\)-\nabla\cdot\(\d_\tau\v{u}^s\)\] = 0$. 

Dislocations arise as singular configurations with non-trivial circulation of $\v{u}$: $\oint \d_iu_j d\ell_i = 2\pi b_j$, where $b_j$ is the $j^\text{th}$ component of the dislocation's Burgers vector, and the integral is taken along any path that encircles the dislocation. Equivalently, we may characterize the $\mu \in \{\tau,x,y\}$ component of the dislocation current with Burgers vector component $i$ of unit length, by:
\begin{align}
J_{\mu,i}^d = \frac{\eps_{\mu\nu\lambda}}{2\pi}\d_\nu\(\d_\lambda u_i\)
\end{align}
where $\eps_{\mu\nu\lambda}$ is the fully antisymmetric unit tensor with space-time indices.

\begin{figure}[t]
\centering
\includegraphics[width=0.7\columnwidth]{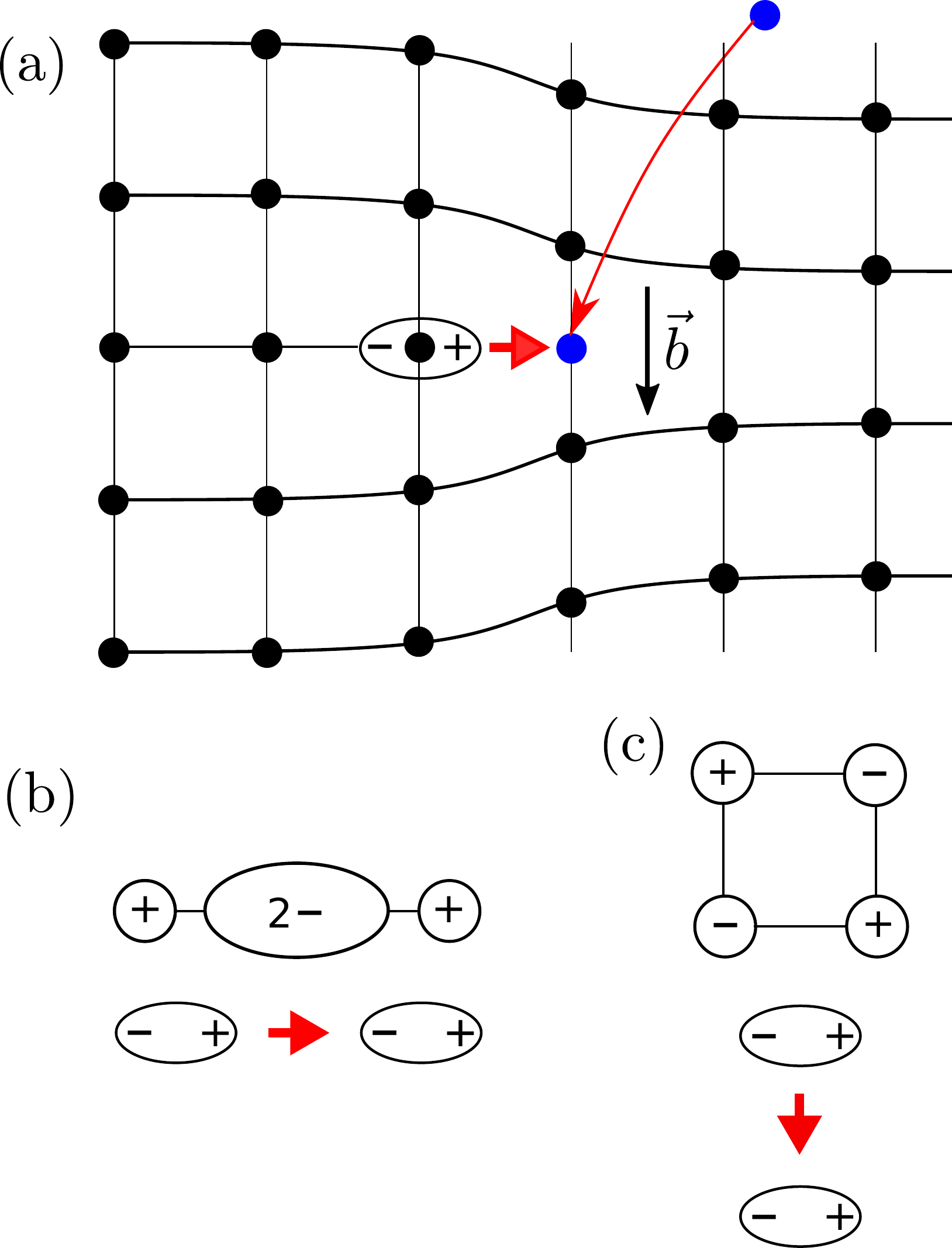}
\caption{{\bf Dislocations motion -- } (a) The dislocation shown with Burgers vector $\vec{b}$ is described by a dipole of gauge charges with dipole moment perpendicular to $\vec{b}$, in the dual gauge theory. (b) \& (c) The climb (glide) motion of the dislocation corresponds to the motion of the dipole parallel (perpendicular) to its dipole moment, and requires the addition of a diagonal (off-diagonal) quadrupole moment. 
 \label{fig:dipole_hop} 
}
\end{figure}

In the presence of a generic dislocation motion, the particle number current is actually not conserved. Instead, one finds:
\begin{align}
\d_\mu j_\mu = 2\pi \rho_0\eps_{ij}J_{i,j}^d
\end{align}
where $\eps_{ij}$ is the antisymmetric unit tensor with spatial indices. Equivalently, to move a dislocation by one lattice spacing perpendicular to its Burgers vector, one must add or remove a unit of charge to the system, as illustrated in Fig.~\ref{fig:dipole_hop}. In a charge-insulating crystal (i.e. one which is not a super-solid or a more exotic compressible quantum liquid state), changing the charge density requires adding energy, and hence each climbing step requires producing gapped charge excitations, preventing the climb motion. Instead, dislocations may only glide along their Burgers vector, and become sub-dimensional fractonic objects that can move only along one-dimensional sub-manifolds of the $2d$ system.


As remarked above, in order to disorder the crystal in a continuous quantum phase transition to a nematic or hexatic phase, one must dynamically condense the dislocations -- a process that is hampered by their symmetry-enforced fractonic nature. The dislocation motion is further inhibited by the long-range interactions between dislocations, mediated by elastic deformations of the crystal and grow logarithmically in the distance between dislocations. In fact, for purely $1d$ systems, it is impossible for particles with such long-range interactions to condense~\cite{giamarchi2004quantum}, but rather they always develop an interaction induced mass or form an incompressible crystal. 

This observation naturally raises the suspicion that it may be fundamentally impossible for the dislocations to condense in a continuous quantum phase transition.
However, the dislocation system is not directly equivalent to a decoupled set of pure $1d$ systems, and requires further analysis. Namely, while each dislocation is confined to move along a $1d$ line, dislocations along different $1d$ lines interact via elastic fluctuations of the crystal. To analyze the resulting system, we develop an effective field theory for dislocations built on previously developed dual theories of elasticity~\cite{beekman2017dual,pretko2018fracton}. In the following section, we briefly review the elegant rank-2 gauge theory formulation of the dual elasticity~\cite{pretko2018fracton}, and then incorporate dislocation fields in a manner consistent with the global number- or charge- conservation of the underlying atoms forming the crystal.

\subsection{Review: Rank-2 dual gauge theory description of elasticity}
To analyze the possibility of a quantum melting transition driven by condensation of fractonic dislocations, we employ a dual rank-2 gauge theory description of the elastic fluctuations developed in Ref.~\cite{pretko2018fracton}, building on previous work of Zaanen \textit{et al}.~\cite{zaanen2004duality}. We will then generalize these theories to include gauge-charged (disclinations) and gauge-dipolar (dislocations) matter. 

The construction of the dual theory starts from the continuum description of elastic fluctuations of the atomic displacement vector field $u_i(x)$ with Lagrangian density:
\begin{align}
\mathcal{L}_\text{el.} = \frac{1}{2} \[ (\d_\tau u_i )^2 + C_{ijkl} \d_i u_j \d_k u_l \]
\end{align}
where $C_{ijkl}$ is the rank-4 elasticity tensor, which is symmetric in arguments $ij$ and $kl$. Following standard duality transformation, we introduce Hubbard-Stratonovich fields $\pi_i$ (the lattice momentum) and $\sigma_{ij}$, where $\sigma_{ij}$ (the symmetric stress-tensor):
\begin{align}
S = \int d\tau d^2x \[ \frac{1}{2} C^{-1}_{ijkl} \sigma_{ij} \sigma_{kl} + \frac{1}{2} \pi_i \pi_i + i \sigma_{ij} \d_i u_j - i \pi_i \d_\tau u_i  \].
\label{eq:elast_hub_str}
\end{align}
Separating the displacement field into: $\v{u} = \v{u}^s+\v{u}^d$, into  smooth (s) part and a singular defect (d) part that accounts for topological defects, and integrating out the smooth field constrains the fields $\pi_i$ and $\sigma_{ij}$ to obey the momentum-conservation continuity equation:
\begin{align}
\d_\tau \pi_i - \d_j \sigma_{ij} = 0.
\label{eq:elast_constraint}
\end{align}
An intriguing analogy to electrodynamics arises upon rewriting stress tensor in terms of rank-2 electric field, and the momentum in terms of its corresponding magnetic field, $b_i$:
\begin{align}
e_{ij} &= \epsilon_{ik} \epsilon_{jl} \tilde{C}^{-1}_{klmn} \sigma_{mn}\nonumber\\
b_i &= \epsilon_{ij} \pi_j, 
\end{align}
where $\eps_{ij}$ is the unit antisymmetric tensor with two spatial indices, and $\tilde{C}_{ijkl} = \epsilon_{ia} \epsilon_{jb} \epsilon_{kc}\epsilon_{ld} C_{abcd}$. In terms of these variables the constraint equation becomes a Faraday-type ``law": 
\begin{align}
\d_\tau b_i + \epsilon_{jk} \tilde{C}_{iklm} \d_j e_{lm} = 0
\label{faraday}
\end{align}

This constraint can be solved, at the expense of introducing a gauge-redundancy, by expressing $e_{ij}$ and $b_i$ in terms of a symmetric rank-2 tensor gauge field $a_{ij}$ and a scalar potential $a_0$, such that Eq.~\ref{faraday} is automatically satisfied:
\begin{align}
e_{ij} &= \tilde{C}^{-1}_{ijkl} \( \d_\tau a_{kl} + \d_k\d_l a_0 \)  \nonumber \\	
b_{i} &= -\eps_{jk}\d_j a_{ki}
\end{align}
The physical fields are invariant under gauge transformations of the form:
\begin{align}
a_{ij} \rightarrow a_{ij}+\d_i\d_j\xi, \hspace{10pt} a_0\rightarrow a_0+\d_\tau\xi
\label{eq:gaugexform}
\end{align}
Writing the action (\ref{eq:elast_hub_str}) in terms of $e_{ij}$ and $b_i$, gives
\begin{align}
S = \int d\tau d^2x \[ \frac{1}{2} e_{ij} \tilde{C}_{ijkl} e_{kl} + \frac{1}{2} b_i b_i -\rho_c a_0 - J_{ij} a_{ij} \]
\end{align}
where $\rho_c$ is the gauge-charge (disclination) density, and $J_{ij}$ is the tensor of disclination currents \cite{pretko2017generalized}, and $a_0$ acts as a Lagrange multiplier to enforce the following Gauss's law constraint:
\begin{align}
\d_i \d_j e_{ij} = \rho_c.
\label{gauss}
\end{align}
Using the definition of the disclination density in the above constraint, one obtains a mapping between the electric field tensor and the strain tensor.
\begin{align}
e_{ij} = \frac{1}{2} \epsilon_{ik} \epsilon_{jl} \( \d_k u_l + \d_l u_k \).
\label{dual_map}
\end{align}

In addition to conservation of the total charge, the Gauss's law constraint (\ref{gauss}) also implies conservation of the total dipole moment, which has the striking consequence that the charges in the theory are immobile fractons~\cite{pretko2017subdimensional,pretko2017generalized,pretko2018fracton}. The result is a dual tensor gauge theory coupled to fractonic matter. The dipoles, in this case, are bound pairs of disclinations, which are dislocations.

The $U(1)$ symmetry enforced fractonic nature of dislocations is captured by the following conservation law, obtained using the Gauss's law (\ref{gauss}).
\begin{align}
\int d^2x \( \rho x^2 - 2 e_{ii} \) = \text{const.}
\label{eq:quadrupole_conservation}
\end{align} 
Using the duality mapping (\ref{dual_map}), one obtains $e_{ii} \approx \Delta n$, where $\Delta n$ is the change in the particle density. Therefore, in a charge insulating crystal, the sum of the diagonal components of the quadrupole moment is conserved, forbidding the motion of dipoles along the direction of their dipole moment, as illustrated in Fig. (\ref{fig:dipole_hop}). The dipole moment is perpendicular to the Burgers vector of the corresponding dislocation, and hence we obtain the 1d particle property of dislocations.

\subsection{Effective theory of dislocations}
Having reviewed the dual higher-rank gauge theory description of phonon fluctuations, we now turn to developing an effective field theory for topological defects in the elastic medium, i.e. to incorporate matter into the dual gauge theory. To this end, we will need to identify the leading relevant operators describing the dynamics of dislocations that are invariant under the rank-2 gauge structure, and carefully consider their transformation properties under the global $U(1)$ symmetry associated with the conserved number of the underlying particles forming the crystal.

\subsubsection{Gauge-charges (disclinations)}
The fundamental gauge-charges of the dual elastic theory are disclinations (defects in the orientational order of the crystal). We can introduce a field, $\psi_c$, for these gauge-charged fields, which transforms under the rank-2 gauge transformations (Eq.~\ref{eq:gaugexform}) as:
\begin{align}
\psi_c(\vec{r},\tau)\rightarrow e^{i\xi(\vec{r},\tau)}\psi_c(\vec{r},\tau)
\end{align}
As previously remarked, the higher-rank gauge structure implies that both the gauge-charge and gauge-dipole moment of the defect matter fields are conserved. Consequently, all gauge invariant operators have vanishing charge and dipole moments. The lowest-order moment operators are those that add gauge-quadrupoles of the matter fields, such as:
\begin{align}
Q_{xy}(\v r,\tau) =& \psi_c(\v{r})\psi^\dagger_c(\v{r}+d\hat{x})\psi_c(\v{r}+d(\hat{x}+\hat{y}))\psi^\dagger_c(\v{r}+d\hat{y})
\times\nonumber\\&\times
e^{-i\int_x^{x+d} dx_1\int_y^{y+d}dy_1 ~a_{xy}(x_1,y_1,\tau)}
\nonumber\\
Q_{xx}(\v{r},\tau) =& \psi_c(\v{r}+d\hat{x})\(\psi^\dagger_c(\v{r})\)^2\psi_c(\v{r}-d\hat{x})
\times\nonumber\\&\times
e^{- i\int_{x}^{x+d}dx_1\int_{x-x_1}^{x+x_1} dx_2 ~a_{xx}(x_2,y,\tau)}
\nonumber\\
Q_{yy}(\v{r},\tau) =& \psi_c(\v{r}+d\hat{y})\(\psi^\dagger_c(\v{r})\)^2\psi_c(\v{r}-d\hat{y})
\times\nonumber\\&\times
e^{-i\int_{y}^{y+d}dy_1\int_{y-y_1}^{y+y_1} dy_2 ~a_{yy}(x,y_2,\tau)}
\label{eq:quadrupole_def}
\end{align}
where $d$ is the lattice spacing.

In a system where the number of particles that form the crystal is a conserved quantity, it is crucial to consider the physical charge quantum numbers of these quadrupolar operators, i.e. their transformation properties under the $U(1)$ symmetry associated with the particle number conservation. 
From the previous section we have seen that that dislocation climb by one lattice spacing requires adding or removing one particle. In the gauge theory, a dislocation is a dipolar composite of two opposite gauge charges that are displaced by distance $\vec{d}$ equal to a lattice spacing. The climb motion, then corresponds to changing moving the gauge-dipole one lattice spacing along its dipole moment. This is precisely the effect of the quadrupolar operators $Q_{xx}$ and $Q_{yy}$ (see Fig.~\ref{fig:dipole_hop}b). Therefore, under  a $U(1)$ rotation that transforms unit-charged operators by $\chi$, these gauge-quadrupole operators must likewise transform:
\begin{align}
Q_{xx/yy} &\underset{{\footnotesize U(1)_{\chi}}}{\longrightarrow} e^{-i\chi}Q_{xx/yy}
\nonumber\\
Q_{xy}&\underset{{\footnotesize U(1)_{\chi}}}{\longrightarrow} Q_{xy}.
\label{eq:quadrupole_u(1)}
\end{align}
Using Eq. (\ref{eq:quadrupole_def}), this transformation then implies that the disclination field transforms as:
\begin{align}
\psi_c(\v{r},\tau) \underset{{\footnotesize U(1)_{\chi}}}{\longrightarrow} e^{i r^2 \chi/2d^2}
\psi_{c}(\v{r},\tau).
\end{align}

Therefore the operators $Q_{xx}$ and $Q_{yy}$ cannot appear alone in the action of any $U(1)$-symmetric theory. Instead, the minimal action for the gauge charges (disclinations) in a number-conserving system is:
\begin{align}
\mathcal{L}_c = \psi_c^\dagger\(i\d_\tau-a_0\)\psi_c+\lambda_{xy}Q_{xy}(\v{r},\tau)+V_c(|\psi_c|^2)
\label{eq:Lc}
\end{align}
where $V_c$ is a potential for the disclinations, and $\lambda_{xy}$ is a non-universal parameter.

\subsubsection{Gauge-dipoles (dislocations)}
Though disclinations are the elementary gauge-charged objects, in a crystalline solid phase, they are not only fully immobile, but also linearly confined~\cite{pretko2017generalized}. Consequently, they do not play a role in the low-energy physics of a continuous melting transition. In contrast, dislocations have a much weaker logarithmic-in-distance interaction (like vortices in a superfluid), which makes them important near the thermal melting transition. In the dual description, dislocations are dipolar composites of the fundamental gauge charges. Let us denote the dislocation field with gauge-dipole moment $\v{d}$ by $\psi_\v{d}$, which corresponds to the field of a dislocation with Burgers vector $b_i=\eps_{ij}d_j$. Since the Burgers vector is quantized to an integer multiple of lattice spacings, the elementary dislocation dipole moment, $|\v{d}|$ is equal to the lattice spacing $d$.

Viewing the dislocation field as tightly bound pair of opposite unit gauge charges separated by unit distance in the $i^\text{th}$ direction, one can deduce its gauge-transformation properties from those of the elementary charge fields, Eq.~\ref{eq:gaugexform}: 
\begin{align}
\psi_{\v{d}}(\v{r},\tau) \rightarrow e^{i\(\xi(\v{r}+\v{d},\tau)-\xi(\v{r},\tau)\)}\psi_{\v{d}}(\v{r},\tau) \approx e^{i\v{d}\cdot \nabla \xi(\v{r},\tau)}\psi_{\v{d}}(\v{r},\tau)
\end{align}
where the last line holds in the continuum limit where we coarse-grain our distance scale on lengths much bigger than the underlying lattice spacing. 

Gauge-invariant operators that move gauge-dipolar particles (dislocations) from spacetime point $x=(r,\tau)\rightarrow x'=(\v{r}',\tau')$ are Wilson lines of the form:
\begin{align}
W_{\v{d}}&(x',x)=
\psi_\v{d}^\dagger(x')e^{i\int_x^{x'} \(d_ia_{ij}d\ell_j + \v{d}\cdot\nabla a_0d\ell_0\)}\psi_{\v{d}}(x)
\end{align}
Taking the limit of infinitesimal length Wilson lines, we see that gauge invariant terms in a continuum dislocation field theory will involve covariant derivatives of the form:
\begin{align}
D_i\psi_{\v{d}} &= \(\d_i-id_ja_{ji}\)\psi_{\v{d}} \nonumber\\
 D_\tau\psi_{\v{d}} &= \(\d_\tau - i\v{d}\cdot\nabla a_0\)\psi_{\v{d}}
\end{align}

To examine the particle-number quantum numbers of the climb operator, consider a $U(1)$ transformation that rotates unit-charged fields by phase $\chi$. We know that the dislocation climb operator adds a particle to the system and will likewise acquire a phase under this $U(1)$ rotation. In the gauge description, this dislocation climb operator, hops a dipolar particle by a unit lattice spacing along its dipole moment, and must therefore transform as:
\begin{align}
\mathcal{O}_{\text{climb},\v{d}}(\v{r}) =& \psi_\v{d}^\dagger(\v{r}+\v{d},\tau)e^{i d_ia_{ij}d_j}\psi_\v{d}^{\vphantom\dagger}(\v{r},\tau) \nonumber\\
 &\underset{{\footnotesize U(1)_{\chi}}}{\longrightarrow} e^{-i\chi}\mathcal{O}_{\text{climb},\v{d}} 
\end{align}

In a charge-conserving system, such charged operators cannot appear on their own in the effective theory, but rather can only enter in charge-neutral composites (perhaps involving other charged fields). Assembling these ingredients, the minimal form of the low-energy effective dislocation field theory compatible with gauge invariance and charge conservation is:
\begin{align}
\mathcal{L}_\text{dis.} &= \sum_{\v{d}} \psi_{\v{d}}^\dagger D_\tau\psi^{\vphantom\dagger}_{\v{d}}
+\frac{1}{2m_d}\left\vert \Pi^{\perp_\v{d}}_{ij}D_j\psi_{\v{d}} \right\vert^2
+ V_d\(|\psi_{\v{d}}|^2\) 
\nonumber\\ &+ \mathcal{L}^{\v{d}}_\text{oct.}
\label{eq:dislocations}
\end{align}
where the $\v{d}$-sum ranges over different elemental lattice vectors, and $\mathcal{L}^{\v{d}}_\text{oct.}$ hops a diagonal gauge-quadrupole along $\v{d}$, or in other words, adds the corresponding gauge-octupole, and:
\begin{align} 
\Pi^{\perp_\v{d}}_{ij}=\(\delta_{ij}-\hat{d}_i\hat{d}_j\)
\end{align}
 is a projection onto the direction perpendicular to the gauge-dipole moment. Crucially, the spatial derivative term omits gradients along the gauge-dipole direction (enforced by the projector $\Pi^{\perp_\v{d}}$), which would correspond to the $U(1)$-forbidden climb motion, and $V_d$ is a (generally non-linear) effective Ginzburg-Landau-type potential for the dislocations. However, $\mathcal{L}^{\v{d}}_\text{oct.}$ corresponds to a pair of dislocation climb motions opposite to each other, or equivalently, the hopping of a crystal particle, and therefore, respects the global $U(1)$ symmetry. On the lattice, it takes the following form.
\begin{align}
\mathcal{L}^{\v{d}}_\text{oct.} = t_{\text{hop}}&\sum_{\v{d^{\prime}}} \mathcal{O}^{\dagger}_{\text{climb},\v{d}}(\v{r}+\v{d^{\prime}}) \mathcal{O}_{\text{climb},\v{d}}(\v{r})  \nonumber + h.c.
\end{align}
In the continuum limit, this gives second derivative terms, as shown below in Eq. \ref{eq:dislocation_theory}. 

For large quantum fluctuations of the atomic positions, dislocations become important and $V_d$ can develop a minimum at a non-zero value of, $|\rho_d|$. It is then useful to decompose the dislocation field into amplitude and phase components
\begin{align}
\psi_{\v{d}} = \sqrt{\rho_{\v{d}}} e^{i\phi_{\v{d}}}.
\end{align}
Integrating out massive fluctuations in the amplitude produces:
\begin{widetext}
\begin{align}
\mathcal{L}_\text{dis.} =& \sum_{\v{d}} \frac{\rho_d}{2}\[\(\d_\tau\phi_{\v{d}}-\v{d}\cdot\nabla a_0\)^2 +\frac{1}{2m_d} \(\Pi^{\perp_\v{d}}_{ij}\(\d_j\phi_{\v{d}}-d_ka_{kj}\)\)^2 + t_{\text{hop}} \sum_{\v{d^{\prime}}} \(\hat{d}_i d^{\prime}_j \d_i \d_j \phi_{\v{d}}-d_i d^{\prime}_k \d_k a_{ij}\hat{d}_j\)^2 \] 
\label{eq:dislocation_theory}
\end{align}
\end{widetext}
where the absence of a linear time-derivative term is guaranteed by the zero net density of dislocations in the system (due to inversion symmetry, which implies a particle-hole symmetry for dislocations). For future reference, we note that the transformation of the dislocation phase field, $\phi_\v{d}$ under $U(1)$ rotations by angle $\chi$ is:
\begin{align}
\phi_\v{d}(\v{r},\tau) \underset{{\footnotesize U(1)_{\chi}}}{\longrightarrow} \phi_{\v{d}}(\v{r},\tau) +\frac{\v{d}\cdot \v{r}}{d^2} \chi
\end{align}

Having constructed effective field theories for the topological defects of the crystal, we can now study the melting phase transitions driven by their condensation. First, we consider the theory of dislocations (\ref{eq:dislocation_theory}) and study their condensation while disclinations remain gapped.
For an ordinary, unconstrained $2+1d$ $XY$ model (where the dislocations have an isotropic kinetic energy), there would be two distinct phases. For small $\rho_d$, the phase fluctuations would be large, producing a gapped state with massive dislocations. Whereas, for large $\rho_d$, the phase fluctuations become stiff, producing long-range phase order. We will now see that the presence of the dynamical
glide-constraint alters this scenario by requiring the onset of a superfluid of vacancies and interstitials in order for the dislocation condensation to occur.

\section{Quantum melting via a super-solid}
\begin{figure}[t]
\centering
\includegraphics[width=\columnwidth]{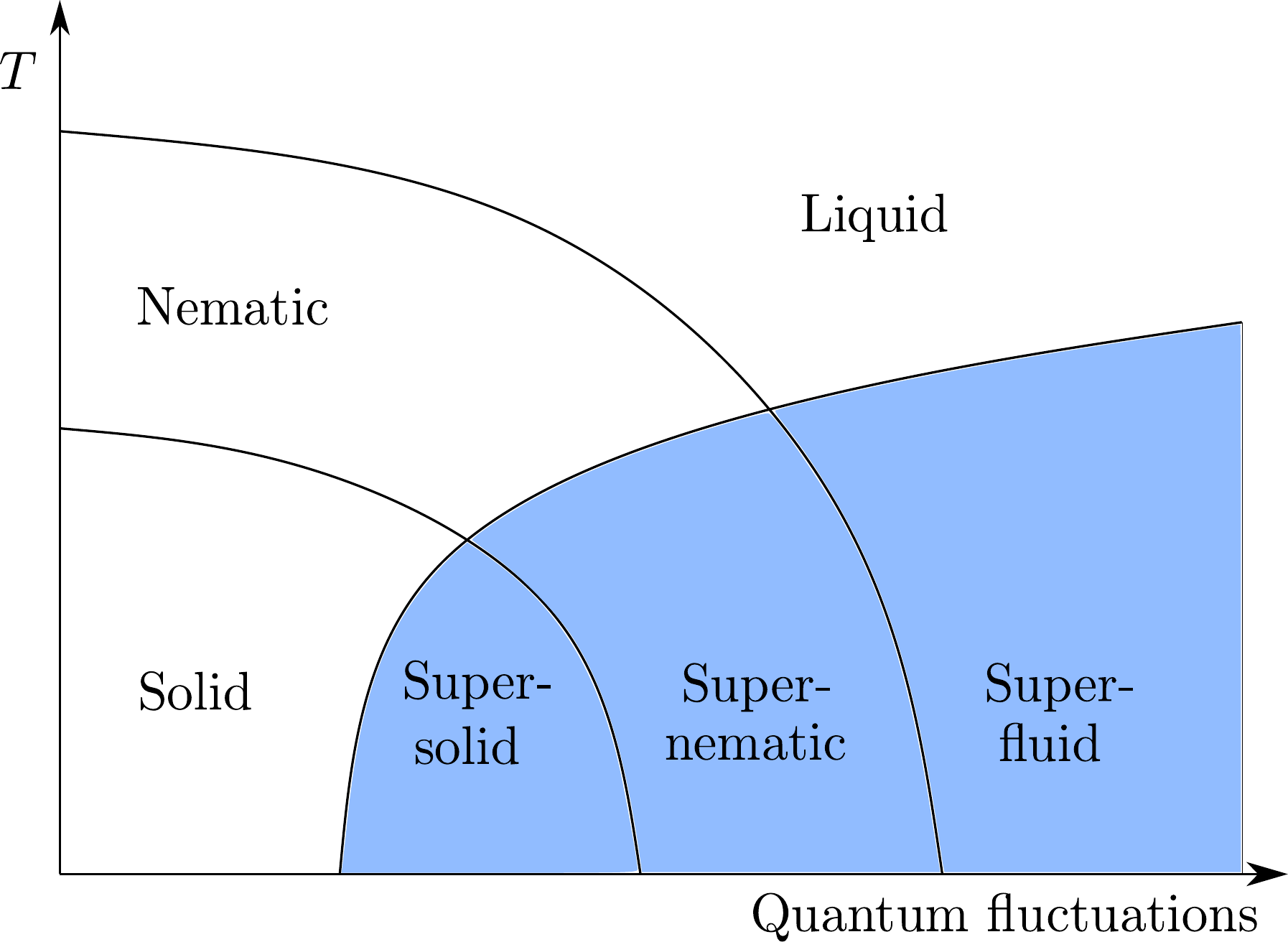}
\caption{Schematic phase diagram - At zero temperature, the fractonic nature of the topological defects of the crystal leads to a departure from the classical Halperin-Nelson thermal melting scenario. We obtain a sequence of disordering phase transitions preceded by the onset of superfluidity of interstitials and vacancies, marked in blue. At finite temperature, thermal fluctuations cause the long-range order of the superfluid to change to quasi-long-range order, and subsequently, to short-range order via a BKT transition at the boundary of the blue region.
 \label{fig:phase_diagram} 
}
\end{figure}

In this section, we show that quantum fluctuations in the crystal can actually favor the formation of an intermediate super-solid phase characterized by co-existing superfluid and crystalline orders. In this super-solid phase, the particle number symmetry is spontaneously broken, freeing the dislocations and enabling them to both glide and climb. This enables a cascade of continuous quantum disordering transitions from solid to super-solid, to super-nematic, and finally to an isotropic superfluid, as shown in the schematic phase diagram in Fig.~\ref{fig:phase_diagram}. 

To construct an effective field theory description for this scenario, we first introduce a Ginzburg-Landau action for the superfluid order parameter $\Psi_\text{sf}$:
\begin{align}
\mathcal{L}_\text{sf} = \Psi_\text{sf}^\dagger i\d_\tau \Psi^{\vphantom\dagger}_\text{sf} + \frac{1}{2m_s}|\nabla \Psi_\text{sf}^{\vphantom\dagger}|^2+\frac{r}{2}|\Psi_\text{sf}|^2+\frac{u}{4!}|\Psi_\text{sf}|^4
\end{align}
where, in a mean-field treatment, $r>0$ would correspond to an insulator, and $r<0$ to an ordered superfluid.

Noting that under $U(1)$ phase rotations, $\Psi_\text{sf}\underset{U(1)_\chi}{\longrightarrow}e^{i\chi}\Psi_\text{sf}$, we see that the minimal gauge invariant and number conserving coupling between the the dislocation climb operators and the superfluid order parameter takes the form:
\begin{align}
\mathcal{L}_\text{sf-dis.} = \gamma \Psi_\text{sf} \sum_{\v{d}}\mathcal{O}_{\text{climb},\v{d}}+h.c.
\end{align}
which describes a process in which the dislocation climbs by removing an atom from the superfluid condensate to conserve the total particle number. We will first examine the back-action of the dislocation fields on the superfluid action, where we will see that the quantum fluctuations of the crystal can actually favor the formation of superfluid order. Subsequently, we will analyze the effective action for dislocations in the resulting super-solid phase.

\subsection{Crystal fluctuation induced super-solidity}
Consider the above theory when the dislocations are gapped, but close to a continuous condensation transition. Integrating out the dislocation fields, produces a renormalized effective potential for the superfluid: $r \rightarrow r_\text{eff} = r - \frac{1}{2}\gamma^2\int d^3x\<\mathcal{O}_{\text{climb},\v{d}}^\dagger(x) \mathcal{O}_{\text{climb},\v{d}}(0)\>$. Since the dislocations are massive, the two point functions of the climb operator decay exponentially producing:
\begin{align}
r_\text{eff} = r - C\gamma^2\int d^3x \frac{e^{-x/\xi_d}}{x^p}
\end{align}
where $C$ is a constant and $p$ is the exponent at the critical point. Close to the critical point, the above integral is $\sim \xi_d^{3-p}$, which diverges at the critical point for $p \leq 3$, and is finite for $p>3$. Hence, for $p \leq 3$, $r$ necessarily gets renormalized to a negative value at a sufficiently large $\xi_d$ and hence the system develops superfluid order before the condensation of the dislocations. In this case, the assumption of a single dislocation condensation transition fails, and instead is preceded by a super-solid phase.

If, instead, $p > 3$, $r$ only gets shifted by a finite constant at the critical point, and in particular, can still remain positive, meaning that the system remains charge insulating. In the dislocation condensed phase (nematic phase), the climb operator $\sim e^{i d_i \d_i \phi_{\v{d}} - d_i a_{ij} d_j}$ is long-range ordered, and therefore, the system has coexisting superfluid order. This leads to the possibility of a continuous phase transition at which the $U(1)$ number conservation symmetry is spontaneously broken and the broken translation symmetry of the crystal is simultaneously restored. Such a scenario requires fine-tuning in the Ginzburg-Landau-Wilson theory of phase transitions, where the generic possibilities are either a first-order phase transition or a pair of separate second order phase transitions corresponding to the two independent symmetries. This suggests that a conventional scenario for the quantum melting transition would have $p<3$.
We note in passing, that such a Landau forbidden transition is a characteristic feature of deconfined critical points, which exhibit emergent fractionalized excitations and gauge degrees of freedom at criticality~\cite{senthil2004deconfined,senthil2004deconfined2}.

Whether such an exotic scenario could arise for quantum crystal melting is intriguing, but we leave the computation of the exponent $p$ for future work (e.g. this question could be addressed by quantum Monte Carlo sampling of dislocation worldlines), and subsequently focus on the more conventional scenario of $p <3$, where crystal melting can continuously occur only via an intermediate superfluid region.
%
%

In such a fluctuation-induced superfluid with large but finite $\xi_d$, the dislocations are still massive and the crystalline order remains intact. Therefore, the resulting phase is a super-solid with both spontaneous breaking of particle number conservation and translation symmetries. Denoting the phase of the ordered superfluid as, $\varphi_s$ ($\Psi_\text{sf}\sim \sqrt{\rho_s}e^{i\varphi_s}$), the effective action of the super-solid phase is:
\begin{align}
\mathcal{L}_\text{ss} &=  \mathcal{L}_\text{el.} + \mathcal{L}_\text{sf} + \mathcal{L}_\text{dis.}+\mathcal{L}_\text{sf-dis.} 
\nonumber\\
\mathcal{L}_\text{el.} &=\frac{1}{2}\(b_i^2+ e_{ij}\tilde{C}_{ijkl}e_{kl}\) \nonumber\\
\mathcal{L}_\text{sf} &= \frac{\rho_s}{2}\(\d_\mu\varphi_s\)^2 \nonumber\\
\mathcal{L}_\text{dis.} & = \frac{\rho_d}{2}\sum_{\v{d}}\(\d_\tau\phi_{\v{d}}-d_i\d_i a_0\)^2 - &
\nonumber\\ &~~ - \frac{\rho_d}{2m_d}\sum_{\v{d}}\cos\(\Pi^{\perp_\v{d}}_{ij}\eps_{il}\(\d_j\phi_{\v{d}}-d_ka_{kj}\)d_l\) \nonumber\\
\mathcal{L}_\text{sf-dis.} &= -\rho_d^\p\sum_\v{d}\cos\(d_i\d_i\phi_{\v{d}}-d_ia_{ij}d_j - \varphi_s\)
\label{eq:ss}
\end{align}
where $\rho_d^\p = \gamma\sqrt{\rho_s}$, and the last term represents the superfluid assisted climb motion of the dislocations, and for simplicity of notation we have used a Lorentz invariant action for the superfluid phase fluctuations. Here, we have not included the particle hopping term $\mathcal{L}^{\v{d}}_\text{oct.}$, in $\mathcal{L}_\text{dis.}$, as it is less relevant compared to the superfluid-climb coupling term.

In the absence of singular vortex configurations in the superfluid phase, we may simply absorb the smooth phase field $\varphi_s$ into the dual rank-2 gauge field, $a_{ij}$ by a suitable gauge transformation. The resulting action has anisotropic, but non-zero kinetic energy for dislocations in both directions -- i.e. the superfluid alleviates the glide constraint and converts dislocations from fractons to ordinary mobile defects. We refer to the dislocations as defects rather than particles, since they still have logarithmic in distance interactions due to the elastic fluctuations, so strictly speaking they are weakly confined.

\subsection{Super-solid to super-nematic transition}
With the symmetry enforced fracton constraint lifted, the crystalline order of the super-solid phase may melt by a continuous quantum phase transitions in which dislocations condense. This transition falls in the $3d$ XY universality class, and produces a super-nematic phase with both coexisting superfluid and orientational-symmetry breaking (but translation symmetry preserving) nematic order. We next examine the low-energy Goldstone mode excitations of the super-nematic phase. The results we obtain agree, where they coincide, with the well-known properties of general nematic superfluids, and have been previously obtained for the rank-1 $U(1)$ gauge field description in Ref.~\cite{zaanen2004duality}. However, we reformulate these results in the rank-2 gauge description.

The super-solid phase contains three types of Goldstone modes: a superfluid phase mode associated with the broken particle-number conservation, and two acoustic phonon branches associated with the broken $x$- and $y$- translation symmetries. In the dual description, these two phonon branches correspond to the two photon branches of the rank-2 gauge field. Whereas, the superfluid phase mode remains essentially unchanged across the super-solid to super-nematic transition, the elastic Goldstone modes are altered. Namely, the super-nematic breaks only a single spatial symmetry (rotation), and hence we expect only a single elastic Goldstone mode, associated with long-wavelength fluctuations of the nematic director. In the dual theory, this arises via a Higgs mechanism in which one of the two photon branches acquires a mass due to coupling with the dislocation (gauge-dipole) condensate.

To see how this arises, we examine the low-energy effective theory of the super-nematic phase, starting from Eq.~\ref{eq:ss}. To simplify the analysis, let us specialize to the square lattice, choose a diagonal elastic tensor, $C_{ij,kl}=c\delta_{ik}\delta_{jl}$, and rescale $a_{ij}$ and $C$ to absorb factors of the gauge-dipole moment $d$, choose $\rho_d=\rho_d^\p$, and pick units in which $m_d=1$. While these choices simplify the analysis, they will not change the general structure of universal features (such as number and character of Goldstone modes). We will choose to work in the axial gauge, where $a_0=0$.  Since vortices in the dislocation phase are suppressed by the dislocation condensate, we can expand the cosine terms in Eq.~\ref{eq:ss} to quadratic order in their arguments. The resulting action reads:
\begin{align}
\mathcal{L}_\text{sn} =& \frac{\rho_s}{2}\(\d_\mu\varphi_s\)^2+\frac{1}{2}\(e_{ij}^2+b_i^2\)+
\nonumber\\
&+\frac{\rho_d}{2}\(\d_\tau \phi_i\)^2+\frac{\rho_d}{2}\(\d_i\phi_j-a_{ij}-\delta_{ij}d^{-1}\varphi_s\)^2
\end{align}

Since the interesting change in the collective mode structure occurs in the elastic sector, let us freeze the superfluid phase by fixing $\varphi_s=0$ and examine the remaining equations of motion for $a$ and $\phi$ (written in terms of a real time coordinate $t=-i\tau$):
\begin{align}
\d_t^2 a_{xx} =& \d_y^2a_{xx}-\d_x\d_ya_{xy}+\rho_d\(\d_x\phi_x-a_{xx}\)
\nonumber\\
\d_t^2 a_{yy} =& \d_x^2a_{yy}-\d_x\d_ya_{xy}+\rho_d\(\d_y\phi_y-a_{yy}\)
\nonumber\\
2 \d_t^2 a_{xy} =& \nabla^2 a_{xy}-\d_x\d_y\(a_{xx}+a_{yy}\)+
\nonumber\\&+\rho_d\(\d_x\phi_y+\d_y\phi_x-2a_{xy}\)
\nonumber\\
\(\d_t^2-\nabla^2\)\phi_i  =& -\d_ja_{ji}
\end{align}

Consider excitations with frequency $\omega$ and wave-vector $\v{q} = q_x\hat{x}$. In the absence of a dislocation condensate, $\rho_d=0$, there would be two propagating photon modes, with either $a_{yy}\neq 0$ (a longitudinal phonon) or $a_{xy}\neq 0$ (a transverse phonon). With non-zero $\rho_d$, the equation of motion for $a_{yy}$ becomes: $\[\omega^2-(q_x^2+\rho_d)\]a_{yy}=0$, i.e. the photon branch corresponding to the longitudinal phonon of the super-solid acquires a Higgs mass. 

In contrast, the off diagonal components of $a_{xy}$ and the dislocation phase $\phi_y$ are coupled, as described by their momentum space equations of motion:
\begin{align}
(\omega^2-q_x^2)\phi_y &=  iq_xa_{xy}
\nonumber\\
(2 \omega^2-q_x^2)a_{xy} &= - \rho_d\(iq_x\phi_y-2a_{xy}\)
\end{align}
Solving these coupled equations, one finds two branches of modes with dispersions: $\omega_{\pm}^2 = \frac{3q_x^2 + 2 \rho_d \pm \( q_x^2+2\rho_d \)}{4}$. The $\omega_+$ branch is gapped for all momenta, but the $\omega_-$ branch contains gapless Goldstone modes with dispersion: $\omega = \frac{1}{\sqrt{2}}|q_x|$


We next verify that this Goldstone mode indeed corresponds to fluctuations of the rotation breaking order. First note, that the fluctuations of $a_{xy}$ and $\phi_y$ do not produce any compression of the crystal atoms. Namely, the local change in density of the crystal is $-\nabla\cdot \v{u}$, in terms of the displacement fields, or equivalently $\sum_i e_{ii}\sim i\omega\(a_{xx}+a_{yy}\)$, which vanishes for this gapless mode. Second, we can compute the distortion of the local bond-angle, $\theta_b = \frac 12 \eps_{ij}\d_iu_j$, associated with this Goldstone mode. To translate this quantity into the dual gauge field variables, we again decompose the displacement field into smooth, and defect parts: $\theta_b = \theta_b^{(s)}+\theta_b^{(d)} = \frac 12 \eps_{ij}\d_i\(u^{(s)}_j+u^{(d)}_j\)$. Note that the elastic contributions to the bond angle can be written as: $\d_\tau \theta^{(s)}_b = \frac12\d_ib_i$, and the dislocation contribution is: $\theta_b^{(d)} =  \int d^2r' \frac{\rho_i^d\eps_{ij} (\v{r}-\v{r})_j}{|\v{r}-\v{r}'|^2}$. Writing the dislocation density, $\rho_i^d = \rho_d\(\d_t\phi_i\)$ (in the axial gauge with $a_0=0$), and inserting the above solution to $\v{q}\sim \hat{x}$ Goldstone mode, one finds that the Fourier components of the bond angle associated with the elastic Goldstone mode are: $\theta_b\sim \frac{i\rho_d}{\omega}a_{xy}$. Together, these observations confirm that the elastic Goldstone mode of the super-nematic phase is indeed a rotational mode as expected.

\subsection{Super-nematic to super-fluid transition}
In the super-nematic phase, the disclinations are no longer immobile fractons. Instead, the disclinations can freely hop without creating further excitations, and have only logarithmic in distance interactions, and thus are weakly confined (since the diagonal components of the rank-2 gauge field, which mediated linear-in-distance interactions in the crystalline phases~\cite{pretko2017generalized} acquire a Higgs mass). As the quantum fluctuations in the nematic order are further increased, these disclination defects can become important at low energies and eventually can condense to destroy the nematic order and restore rotational invariance. 

To describe the disclination condensation process in the effective dual elastic theory, we should consider the disclination field $\psi_c$ with action Eq.~\ref{eq:Lc}.
Note that disclination/anti-disclination pairs displaced by a single lattice spacing are gauge-dipoles, and can freely convert into dislocations, which have the same gauge-dipolar structure. This coupling is described by a term:
\begin{align}
\mathcal{L}_{c-d}&[\psi_c,\psi_d] 
\nonumber\\&= 
-\gamma_{c-d}\sum_\v{d}\psi_{\v{d}}(\v{r},\tau)\psi_c^\dagger(\v{r}+\v{d},\tau)\psi_c^{\vphantom\dagger}(\v{r},\tau)+h.c.
\end{align}

In the super-nematic phase, the dislocations are condensed, i.e. effectively $\<\psi_d\>\neq 0$. To describe this, it is useful to introduce phase variables: $\psi_c=\sqrt{\rho_c} e^{i\phi_c}$ and $\psi_\v{d} = \sqrt{\rho_d} e^{i\phi_\v{d}}$. In the dislocation condensed phase, $\phi_\v{d}$ is approximately constant, which we can set to $0$, and the gauge charge-dipole coupling term becomes effectively $\mathcal{L}_{c-d}\approx -\gamma_{c-d}\sqrt{\rho_d\rho_c}\sum_\v{d}\cos\(\v{d}\cdot\nabla \phi_c\)$. This takes the form of an effective ```hopping"-type kinetic energy for the disclinations. Hence we see that the disclinations, which were immobile fractons in the crystal, may now hop by absorbing dislocations from the dislocation condensate.

\begin{widetext}
To analyze the properties of this in the dual gauge description, let us write down the full effective action for elastic fluctuations, defects, and superfluid degrees of freedom in the phase-only approximation:
\begin{align}
\mathcal{L}_c &\approx \frac{\rho_c}{2}\(\d_\tau\phi_c-a_0\)^2+\frac{\rho_c}{2m_c}\(\d_i\d_j\phi_c-a_{ij}-d^{-2}\varphi_s\delta_{ij}\)^2
\nonumber\\
\mathcal{L}_{\text{dis.}} &\approx \sum_{\v{d}} \frac{\rho_d}{2}\(\d_\tau\phi_\v{d}-\v{d}\cdot\nabla a_0\)^2+\frac{\rho_d}{2m_d}\[\(\Pi^{\perp_\v{d}}_{ij}\(\d_j\phi_{\v{d}}-d_ka_{kj}\)\)^2+\(\v{\hat{d}}\cdot\nabla \phi_\v{d}-\hat{d_i}a_{ij}d_j-d^{-1}\varphi_s\)^2\]
\nonumber\\
\mathcal{L}_\text{ss.}&= \frac{1}{2}b^2+\frac{1}{2}e\tilde C e+\frac{\rho_s}{2}\(\d_\mu\varphi_s\)^2
\end{align}
Writing the dislocation-disclination coupling as $\mathcal{L}_{c-d}\sim -\sum_\v{d}\cos\(\phi_\v{d}-\v{d}\cdot\nabla \phi_c\)$, we see that in the phase where both dislocations and disclinations are condensed, these terms lock $\v{d}\cdot\nabla\phi_c = \phi_\v{d}$, forcing the two components of the dislocation phase fields $\phi_i$ to be gradients of a single scalar $\phi_c$.
\end{widetext}
In this case, we can perform a gauge transformation $a_{ij}\rightarrow a_{ij}+\d_i\d_j\phi_c$, and $a_0\rightarrow a_0+\d_{\t} \phi_c$ to remove the disclination phase field from the action, analogous to the familiar unitary gauge for superconductors. The resulting action in this unitary gauge is:
\begin{align}
\mathcal{L}_c+\mathcal{L}_d\approx 
\frac{\rho_c}{2}a_0^2+\frac{\rho_cm_c^{-1}+\rho_dm_d^{-1}}{2}\(a_{ij}-d^{-2}\varphi_s\delta_{ij}\)^2
\end{align}
so that all the components of the rank-2 gauge structure acquire Higgs masses that lock $a_0=a_{xy}=0$, and $a_{xx}=a_{yy}=\varphi_s d^{-2}$. The only remaining excitation is the superfluid phase mode, signaling that the crystalline order has been completely disordered by the defect condensation, restoring full translation and rotation symmetry, and resulting in an isotropic superfluid. We summarize the sequence of quantum melting phase transitions in Table~\ref{table:summary}.

\begin{table*}[t]
\begin{tabular}{ m{3.5cm}  m{3.5cm} m{4.5cm}  m{4cm}  } 
\hhline{====}
\textbf{Phase} &\textbf{Gapless modes} &\textbf{Dislocations} &\textbf{Disclinations} \\
\hline
Solid &$a_{xy}, a_{yy}$ &$1d$ fractons, gapped &$0d$ fractons, gapped \\ 

Super-solid &$a_{xy}, a_{yy}, \varphi_s$ &unconstrained, gapped &$0d$ fractons, gapped \\ 

Super-nematic &$a_{xy}, \varphi_s$ &unconstrained, condensed &unconstrained, gapped \\ 

Super-fluid &$\varphi_s$ &unconstrained, condensed &unconstrained, condensed \\ \hhline{====}
\end{tabular}
\caption{Summary of the zero-temperature phases, gapless modes propagating along the $x$-direction, and the nature of the topological defects. The photon modes $a_{yy}$ and $a_{xy}$ acquire Higgs masses when the dislocations and the disclinations are condensed respectively, and $\varphi_s$ is the Goldstone mode corresponding to the condensation of the bosons that constitute the lattice. }
\label{table:summary}
\end{table*}

\subsection{Finite temperature cross-over} \label{subsec:finite_temperature}
At zero temperature, we have seen that the melting of a crystal to a nematic phase is accompanied by a superfluid of vacancies and interstitials, where the dislocations lose their sub-dimensional property. However, one must recover the classical Halperin-Nelson scenario at finite temperature where there is a continuous phase transition from the crystal to the nematic phase. To examine how this works, we must consider the effects of thermal fluctuations.

A detailed discussion of the finite temperature physics of fractonic matter was presented in Ref.~\cite{pretko2017finite}. The basic point is that fractons are only prevented from moving at zero temperature, where each forbidden move requires exciting additional gapped fracton excitations. However, non-zero temperature excites a finite density of fractons, which can be absorbed or emitted to allow the forbidden motion. 

Here, a similar scenario holds for symmetry-enforced fractons, with the distinction that it is a thermally excited gas of mobile charged particles that liberate the dislocations at non-zero temperature, as opposed to a thermal self-liberation of inherently fractonic particles. 
Specifically, for energy gap $\Delta$ to charged-particle excitations of the crystal, non-zero temperature will produce a thermal gas of non-crystalline particles of density $\rho_T\sim e^{-\Delta/T}$. Dislocations can then climb by absorbing or emitting particles into this thermal gas. Since thermally excited particles are required to assist these climb-direction ``hops" processes, their amplitude will be proportional to $\rho_T$, and hence will also display the same activated temperature dependence. A second distinction from the $3d$ thermal liberation of intrinsic fractons described in Ref.~\cite{pretko2017finite}, is that thermal screening of $3d$ fractons results in weaker power-law interactions, and whereas the dislocations continue to exhibit logarithmic confinement, up until a BKT transition, as described by the Halperin-Nelson theory.

We note that, for temperatures $T\ll \Delta$, there will be very few excited charged particles to assist the dislocation climb, and the dynamics of the dislocations will be highly anisotropic, such that the climb motion is thermally frozen out in the limit of $T\rightarrow 0$. For $T\gtrsim \Delta$, the anisotropy is less pronounced, and the classical Halperin-Nelson scenario can take over.

Finally, a possible alternative is that of a single first order phase transition between the solid and liquid (at non-zero temperature) or between the solid and superfluid at zero temperature.

\section{Vortex lattice melting} \label{sec:vortex_lattice}
The above discussion focused on the case where the underlying crystal arose from bosonic atoms with a conserved number. However, these principles have implications for other types of crystals. As an example, we next consider the quantum melting of a $2d$ vortex lattice of a superfluid or superconductor, where the objects forming the crystal are collective topological defects of a different order. This scenario is closely related to the one discussed above through boson-vortex duality, though there will be some phenomenological differences due to the absence of time-reversal symmetry due to the magnetic field required to produce the vortex lattice state.

Starting from a superfluid (superconductor) state, a vortex lattice can be induced by externally breaking time-reversal symmetry by applying a net rotation (or external magnetic field) respectively. Again denoting the displacement field of the vortex positions by $\vec{u}$, the low-energy effective field theory for this state can be constructed by performing standard boson-vortex duality, in terms of a vortex field, $\psi_v$, minimally coupled to an emergent $U(1)$ gauge field, $\alpha_\mu$ whose flux is $2\pi$ times the density of the particles forming the superfluid. The resulting effective theory is:
\begin{align}
\mathcal{L} &= \mathcal{L}_v[\psi_v,\alpha]+\mathcal{L}_\text{el.}[u]+\mathcal{L}_\alpha[\alpha,A]
\nonumber\\
\mathcal{L}_v & = \bar\psi_v \(-i\d_\tau-\mu-\alpha_0\)\psi_v +\frac{1}{2m_v}|\(\nabla-i\vec\alpha\)\psi_v|^2 +\nonumber \\&  \hspace{10pt} +V(|\psi_v|^2)
\nonumber\\
\mathcal{L}_\text{el.} &= \frac{i}{2}\eps_{ij}u_i\d_\tau u_j + \frac{1}{2}C_{ijkl}\d_iu_j\d_ku_l + \alpha_0\nabla\cdot \v{u}+\v{\alpha}\cdot \d_\tau \v{u}
\nonumber\\
\mathcal{L}_\alpha &= \frac{\(\eps_{\mu\nu\lambda}\d_\nu\alpha_\lambda\)^2}{4\kappa^2}+\frac{i\eps_{\mu\nu\lambda}\alpha_\mu\d_\nu A_\lambda}{2\pi}
\end{align}
where $\mu$ is a chemical potential for the vortices, produced by the external time-reversal breaking field that induced the vortex lattice, $m_v$ is the effective mass of the vortex excitations, and $V(\dots)$ is an effective potential for the vortex excitations (additional vortices beyond those already present in the vortex lattice are gapped excitations of the vortex liquid). In the last line, $\kappa$ is the gauge-coupling of the emergent gauge field, $\alpha$ (here, for simplicity we have written a Lorentz invariant form, though such a symmetry will not naturally be present in the theory, deviations from this form will not alter the subsequent discussion). For vortex lattices in a superfluid or superconductor respectively, $A$ is either: i) an external (i.e. non-dynamical or ``background") electromagnetic potential field (for the superfluid) or ii) the vector potential of a fluctuating magnetic field (for the superconductor). 


\subsection{Phonons and dual elasticity theory}
The absence of time-reversal symmetry dramatically alters the phonon spectrum of the vortex solid compared to an ordinary crystal. Formally, this enables the single time derivative term in $\mathcal{L}_\text{el.}$, which is ordinarily forbidden, but now dominates the usual $\(\d_\tau u\)^2$ form at low energies. Consequently, instead of distinct longitudinal and transverse branches of acoustic phonons, the vortex lattice exhibits a single phonon mode that is a mixture of compression and rotation, and which has a non-relativistic $\omega\sim q^2$ dispersion. We note that other magneto-elastic systems, for example phonons of a skyrmion crystal, are described by an identical magneto-elastic theory~\cite{zang2011skyrmions}:
\begin{align}
\mathcal{L}_\text{m.el.}[u] = \frac{i}{2}\eps_{ij}u_i\d_\tau u_j + \frac{1}{2}C_{ijkl}\d_iu_j\d_ku_l 
\end{align}

Starting from the time-reversal asymmetric elasticity action, we can derive a dual field theory for the vortex crystal phonons, following analogous steps to those of Ref.~\cite{pretko2018fracton}, which has been independently obtained by Zhai \textit{et al}.~\cite{zhaivortex}. We first introduce Hubbard-Stratonovich fields $\pi_i$ and $\sigma_{ij}$ to produce an action that is linear in $u$:
\begin{align}
\mathcal{L}_\text{m.el.}[u,\pi,\sigma] =& \frac{i}{2}\eps_{ij}\pi_i\d_\tau \pi_j + i\pi_i\d_\tau u_i + 
\nonumber\\ 
&+\frac{1}{2}\sigma_{ij}C^{-1}_{ijkl}\sigma_{kl}-i\sigma_{ij}\d_iu_j
\end{align}
The first term marks the chief departure from the time-reversal invariant action. In particular, varying with respect to $\pi_i$, we see that $\pi_i\approx \eps_{ij}u_j$. This implies that the original translation symmetry of the underlying elastic action: $u_i\rightarrow u_i+\delta_i$, for any constant vector $\delta_i$, will require invariance under $\pi_i\rightarrow \pi_i+\eps_{ij}\delta_j$.

Next, decomposing the displacement field into smooth (s) and singular defect (d) parts: $u= u^s+u^d$, and integrating out the smooth fluctuations enforces the constraint of Eq.~\ref{eq:elast_constraint} of the main-text, which can again be solved by introducing a rank-2 gauge structure identical to that explained in Ref.~\cite{pretko2018fracton} and above.

The resulting dual elasticity action reads:
\begin{align}
\mathcal{L}_\text{dual}[a] = \frac{i}{2}b_i\eps_{ij}\d_\tau b_j + \frac{1}{2}e_{ij}\tilde{C}_{ijkl}e_{kl}.
\label{appeq:Ldual}
\end{align}
The $b^2$ term, present in the time-reversal symmetric case is notably absent. Instead the dual magnetic field term contains a linear time-derivative of $b$, such that the dual-photon $a_{ij}$ exhibits the $\omega \sim q^2$ dispersion expected of the magneto-phonons. In fact, the above-noted $\pi_i\rightarrow \pi_i+\eps_{ij}\delta_j$ invariance of the action, which was a consequence of the underlying translation symmetry, manifests as a $b_i\rightarrow b_i -\delta_i$ invariance of the dual action. Therefore, translational invariance forbids any terms involving $b$ without derivatives, preventing the nominally more relevant $b^2$ terms from appearing, and enforcing the correct quadratic phonon (dual-photon) dispersion.

\subsection{Constraints on melting transitions}
The key feature of the above description is that the vortex number becomes the charge of the dual gauge field, $\alpha_\mu$, and is hence conserved. Physically, this vortex number conservation arises since no local fluctuation of the superconductor can change the global vorticity. In the dual elasticity theory, this again means that the dislocation climb terms must couple to a field with compensating gauge-charge under $\alpha$, such as:
\begin{align}
\mathcal{L}_\text{climb} = \Gamma\(e^{i \(d\d_x\phi_x-d^2a_{xx}\)}+ e^{i \(d\d_y \phi_y-d^2a_{yy}\)}\)\psi_v +\text{h.c.}
\end{align}
where $\Gamma$ is some non-universal coupling constant.
Hence, as with the previous analysis, it is not possible to directly condense the dislocations without condensing the vortices, to destroy the underlying superfluid or superconducting state. 

Simply condensing the vortices on top of the background vortex lattice would produce an insulating state, still with translation symmetry breaking crystalline order. The dislocations of this insulating crystal state that descends from the phase-disordered vortex lattice, are no longer constrained to glide by the vortex-number conservation. However, this does not mean that they are free to move. To see this, note that, the flux of the dual gauge field $\alpha$ is locked to the particle number: $\nabla\times \alpha = 2\pi \rho$, so that the vortices are forced to admit a finite gauge-flux density. To accommodate this gauge-flux, the vortex condensate cannot be spatially homogeneous, but must itself have a lattice of dual-vortex defects. In the original particle language this dual-vortex lattice, is simply a crystal of the particles. As we have seen in the previous section, dislocations of this crystal are symmetry enforced fractons that cannot climb, due to particle number conservation. Hence, by condensing the vortices, we have gone from a vortex lattice whose dislocations cannot climb due to the topological vortex-number conservation, to an ordinary insulating crystal whose dislocations cannot climb due to particle-number conservation.


\section{Charge density wave melting in a metal}
Before concluding, we turn our attention to electronic $2d$ crystals such as charge density waves (CDW), and stripe phases of high-temperature cuprate superconductors~\cite{kivelson2003stripesreview,vojta2009lattice,keimer2015quantum}, or other strongly correlated compounds~\cite{xi2015strongly}; the related situation where the $2d$ crystal forms in an electronic system due to interactions.

Electron-electron interactions in a metal can produce spontaneous charge-density wave (CDW) order at wave-vector $Q$ that may or may not be commensurate with the underlying ionic lattice, and which are sometimes accompanied by spin-density wave (SDW) order. A controlled theory of the onset CDW or SDW remains challenging, since the critical fluctuations of the DW order are strongly coupled to the continuum of gapless particle-hole excitations of the Fermi-surface~\cite{abanov2000spin, abanov2004anomalous, metlitski2010sdw}. We will not attempt to address this challenging situation, and instead examine the constraints placed by the symmetry-enforced fractonic nature of its dislocation defects, and comment on open issues for future work.

Suppose we can tune a parameter in the system, such as doping or pressure, that tends to destroy the CDW order. Then, as the quantum fluctuations in the CDW order increase, they will favor dislocation motion. However, as the glide constraint dictates, dislocations cannot climb without absorbing electrons from the Fermi surface into the CDW order. 

Naively, the minimal such process would be for a dislocation to climb enough to absorb a single electron from near the Fermi-surface. This process would be forbidden in a time-reversal invariant system, since the electron carries a spin-1/2. It could conceivably occur in a spin-density wave system where time-reversal symmetry is absent. However, in the above field theory formulation the dislocation climb operator is a bosonic field, and hence cannot couple directly to a single electron creation or annihilation operator (which would correspond to an unphysical non-local conversion of fermions to bosons). It remains an open question whether this statistical obstruction is fundamental, or whether it is possible to modify the dislocation description in such a way to convert the climb operator to a fermion object. 

To simplify our discussion, we will instead focus on a time-reversal invariant CDW state, in which case dislocations can only climb by adding or removing spin-singlet \emph{pairs} of electrons. We can, then re-visit the above analysis of the previous sections, but replacing the role of the bosonic superfluid order parameter, $\Psi_\text{sf}$ by a Cooper pairing field $\Psi_\text{Cp}$ which is a charge-2e, spin-singlet operator, producing a coupling of the form:
\begin{align}
\mathcal{L}_\text{climb} = \Gamma \(e^{2i \(d\d_x\phi_x-d^2a_{xx}\)}+ e^{2i \(d\d_y \phi_y-d^2a_{yy}\)}\)\Psi_\text{Cp}+\text{c.c.}
\nonumber\\
\Psi_\text{Cp}(\vec{r},\tau) = \sum_k e^{iq\cdot r}\Delta_{k,q}(\tau) c_{k+q/2,\up}c_{-k+q/2,\down}
\nonumber\\
\end{align}
Here, the pairing parameter $\Delta_{k,q}$ can encode any symmetry allowed pairing (e.g. s-wave, d-wave, etc...). 

Then, by analogy to the bosonic quantum melting story presented above, we see that the quantum fluctuations in the CDW metal favor the formation of a superconducting electron-pair condensate in order to alleviate the glide constraint and allow the crystal defects to climb, following which, they could condense to melt the CDW metal to a nematic metal (note that more detailed microscopic input is needed to decide which pairing symmetry would be most favorable).~\footnote{Landau damping of the dislocation currents by the Fermi-surface is less relevant than the long-range phonon-mediated dislocation interactions, as further discussed in Appendix~\ref{app:stripe}, and hence is not expected to effect our analysis of the bosonic crystal.} This scenario is reminiscent of the notion of competing orders~\cite{keimer2015quantum} in high-temperature superconductors, and the symmetry-enforced fracton concepts could potentially provide a useful new perspective on these complex materials.


\section{Discussion}
To summarize, we have introduced the notion of symmetry enforced fractonicity, where in the presence of a global symmetry, certain point- particles or defects cannot move along certain directions without exciting gapped excitations that carry non-trivial quantum numbers under that symmetry. We have shown that dislocations of $2+1d$ insulating solids are a simple example of this concept. Moreover, our analysis shows that the symmetry enforced sub-dimensional nature of dislocations dramatically alters the critical properties of quantum melting transitions from what was conjectured in previous literature\cite{zaanen2004duality,beekman2017dual}.

Namely, by developing a dual higher-rank gauge theory description of these symmetry enforced fractonic defects, we have shown that they can condense to drive a continuous quantum phase transition from a crystal to a nematic phase only when the sub-dimensional constraint on them is lifted through the onset of a superfluid of vacancies and interstitials. In retrospect, hints of this result were evident from more general considerations. Namely, the symmetry-enforced fractonic nature of dislocations mean that any dislocation condensate must also spontaneously break the $U(1)$ symmetry associated with number conservation. Therefore, a direct transition from crystal to nematic must simultaneously restore translation symmetry and break number conservation. In a conventional Ginzburg-Landau-Wilson framework, it is generically not possible to go between two phases with completely different symmetry breaking patterns without fine-tuning. Instead, the transition is generically first order, or splits into a pair of separate second order transitions. In other contexts, more exotic deconfined critical points can arise, in which such a Landau-forbidden change of symmetry can occur in a direct continuous quantum phase transition \cite{senthil2004deconfined}. It would be intriguing to consider whether a possible deconfined critical scenario could apply to quantum melting of a solid to a super-nematic. However, at this point we do not have a concrete scenario for such an exotic transition. Even if such a phase transition existed, its critical exponents would be dramatically different than the conventional $2+1d$ XY universality class conjectured in previous works~\cite{zaanen2004duality,beekman2017dual}.

It would also be potentially interesting to investigate the applicability of the concept of global symmetry-enforced fractonicity in other contexts, including $3+1d$ crystals~\cite{pai2018fractoniclines}, or true fracton ``topological phases" in which the sub-dimensional objects are deconfined particles with possible anyonic behavior and exponential in system-size ground-state degeneracy~\cite{chamon2005glassiness,bravyi2011topological,haah2011code,vijay2015xcube}, rather than confined defects of a symmetry-breaking phase.

\vspace{12pt}\noindent{\it Acknowledgements -- We thank M. Pretko and P.T. Dumitrescu for insightful conversations. This work was supported by NSF DMR-1653007.}

\vspace{12pt}\noindent{\it Note -- During the course of completing this manuscript, we became aware of a related work by M. Pretko and L. Radzihovsky~\cite{pretko2018symmetry} on closely related subject matter.}

\appendix

\section{Coupling of stripe dislocations to a Fermi-surface of electronic excitations}\label{app:stripe}
In this appendix, we consider possible couplings of phonons and dislocations of an incommensurate CDW with the reconstructed Fermi-surface. We will find that the crystal excitations have only irrelevant couplings to the low-energy particle-hole continuum of electronic excitations, and hence can be neglected in the analysis of dislocation condensation. 

Since the fermionic fluctuations have vanishing rank-2 gauge charge, they cannot couple directly to the dual gauge field, $a_{ij}$, but rather only to its field strengths, $b_i$ and $e_{ij}$. These couplings contain derivatives that make them irrelevant at low-energies. Intuitively, this reflects the familiar fact that the dual photons described Goldstone modes of broken translation symmetry, which quite generally decouple from other low-energy excitations~\cite{watanabe2014criterion}.

The low energy effective field theory of the Fermi-surface is:
\begin{align}
\mathcal{L} = \int d\varphi ~\psi^\dagger_{\varphi}\(-i\v{v}_F(\varphi)\cdot \nabla\)\psi^{\vphantom\dagger}_\varphi
\end{align}
where $\varphi$ denotes the angle in momentum space (here we assume a single sheet for the reconstructed Fermi-surface for notational simplicity), $v_F$ is the Fermi velocity normal to the Fermi-surface at angle $\varphi$, and $\psi_\varphi(r) = \int_{k_F(\varphi)-\Lambda}^{k_F(\varphi)+\Lambda} dk(\varphi) e^{-i\v{k}_F(\varphi)\cdot \v{r}}\psi_{\v{k}(\varphi)}$ is the low-energy fermion field near the Fermi surface at angle $\varphi$, where $\Lambda$ is a UV cutoff that is less than the Fermi wavelength.

The low-energy modes of the Fermi surface are shape fluctuations described by operators:
\begin{align}
M[f] = \int d\varphi~ f(\varphi) \psi^\dagger_\varphi\psi^{\vphantom\dagger}_\varphi
\label{appeq:fsshape}
\end{align}
where $f$ is some real-valued function of the Fermi-surface angle, $\varphi$. The minimal coupling between the dislocation currents and the low-energy modes of the Fermi-surface would be of the form: $\lambda J_{\mu,i}^d M[f]$ where $M[f]$ has the right spatial symmetries to couple to the $J_{\mu,i}$. In a Hertz-Millis type analysis~\cite{hertz1976quantumcritical,millis1993quantumcritical}, integrating out the electronic fluctuations would produce a Landau damping term $\sim \lambda^2\frac{|\omega|}{q}|J_{\mu,i}^d(q,\omega)|^2$. If we write $J_{\mu,i}^d \sim \eps_{i\mu\nu} \d_\nu \theta_i$ this term only produces interactions of the form: $\sim \lambda^2 \omega q|\theta_i(\omega,q)|^2$. These interactions are less relevant than the long-range dislocation interactions induced by the gauge coupling $g$, and are not expected to alter the analysis presented in the main text for a bosonic crystal.

\bibliography{QCrystalMeltingbib}

\end{document}